\pdfoutput=1
\documentclass[aps,prl,twocolumn,showpacs,superscriptaddress]{revtex4-1}

\def\simgt{\mathrel{\lower .3ex \rlap{$\sim$}\raise .5ex \hbox{$>$}}}

\newcommand{\ket}[1]{ |{#1} \rangle }

\newcommand{\half}{\frac{1}{2}}

\newcommand{\units}[1]{\ensuremath{\mathrm{#1}}}
\newcommand{\amount}[2]{\ensuremath{#1\:\units{#2}}}
\newcommand{\sym}[2]{\ensuremath{#1_{\mathrm{#2}}}}

\newcommand{\Tm}{ T$_-$ }
\newcommand{\Tz}{ T$_0$ }

\usepackage[pdftex]{graphicx}
\usepackage[pdftex]{epsfig}
\usepackage{epstopdf}
\usepackage{color}
\usepackage{amssymb,amsmath}
\usepackage{graphicx}
\usepackage{dcolumn}
\usepackage{multirow}
\usepackage{hyperref}

\begin{document}

\def\simlt{\mathrel{\lower .3ex \rlap{$\sim$}\raise .5ex \hbox{$<$}}}
\def\half{ \frac {1}{2} }

\title{A fast ``hybrid'' silicon double quantum dot qubit}

\author{Zhan Shi}
\author{C. B. Simmons}
\author{J. R. Prance}
\author{John King Gamble}
\author{Teck Seng Koh}
\author{Yun-Pil Shim}
\affiliation{Department of Physics, University of Wisconsin-Madison, Madison, WI 53706, USA}
\author{Xuedong Hu}
\affiliation{Department of Physics, University at Buffalo, SUNY, Buffalo, NY 14260}
\author{D. E. Savage}
\author{M. G. Lagally}
\author{M. A. Eriksson}
\author{Mark Friesen}
\author{S. N. Coppersmith}
\affiliation{Department of Physics, University of Wisconsin-Madison, Madison, WI 53706, USA}

\pacs{03.67.Lx,73.63.Kv,85.35.Be}

\begin{abstract}
We propose a quantum dot qubit architecture that has an attractive combination of speed and fabrication simplicity.  It consists of a double quantum dot with one electron in one dot and two electrons in the other. The qubit itself is a set of two states with total spin quantum numbers $S^2=3/4$ ($S=\half$) and $S_z = -\half$, with the two different states being singlet and triplet in the doubly occupied dot.
The architecture is relatively simple to fabricate, a universal set of fast operations can be implemented electrically, and the system has potentially long decoherence times.  These are all extremely attractive properties for use in quantum information processing devices.

\end{abstract}

\maketitle

Using electrically-gated quantum dots in semiconductor heterostructures
to make qubits for quantum information 
processing~\cite{Loss:1998p120,Vrijen:2000p1643}
is attractive 
because of the potential
for excellent manipulability, scalability, and for integration with classical electronics.
Tremendous progress towards the development of working electrically-gated quantum
dot qubits
has been made over the past decade, and
single-qubit operations have been demonstrated for logical qubits implemented in
single~\cite{Koppens:2006p766}, double~\cite{Petta:2005p2180}, 
and triple~\cite{Laird:2010p1985} quantum dots in GaAs heterostructures.
However, even with sophisticated pulse sequences that lead to 
coherence times up to 200 $\mu s$~\cite{Bluhm:2011p109},
the important figure of merit, the number of gate operations that
can be performed within the qubit
coherence time~\cite{DiVincenzo:1995p2698,Preskill:1998p469,Fisher:2003p1808}, 
needs to be
improved significantly for quantum dot qubits to become useful.
Moreover, it is highly desirable that a given implementation be as simple as possible.

In this paper, we present a relatively simple
double-dot qubit architecture
in which a
universal set of fast gate operations can be implemented.
Each qubit consists of a double quantum dot with
two electrons in one dot and one electron in the other.
The qubit itself is the set of two low-lying electronic states with total spin
quantum numbers $S=\half$ (square of the total spin of 
$\frac{3}{4}\hbar^2$) and $S_z = {-\half}$ (z-component of total spin of $- \frac{\hbar}{2}$). 
These states form
a decoherence-free subspace~\cite{Lidar:1998p2594} that
is insensitive to long-wavelength magnetic flux noise.
Additionally, spin-conserving decoherence processes such as charge noise do not induce transitions that go outside of the subspace of an individual qubit.
All of the
gate operations are implemented using purely electrical manipulations,
enabling much faster gates than
using either ac magnetic fields~\cite{Loss:1998p120,Koppens:2006p766}
or inhomogeneous dc magnetic 
fields~\cite{Levy:2002p1446,Petta:2005p2180,Bluhm:2011p109}.
The qubit has the same symmetries in spin space as the triple-dot
qubit proposed by DiVincenzo et al.~\cite{DiVincenzo:2000p1642}, but
is simpler to fabricate because it requires a double dot instead of a triple dot.
The hybrid qubit proposed here also has significant advantages over the three-dot qubit
for implementing multi-qubit 
operations: two hybrid qubits made of four dots in a linear array have higher
effective connectivity than the similar linear array of dots considered in
Ref.~\cite{DiVincenzo:2000p1642}.  This increased effective connectivity reduces the number of nontrivial
gate manipulations required to implement two-qubit gates.

We present evidence that implementing this qubit in silicon is feasible.
The development of qubits in silicon has attracted substantial
interest~\cite{Kane:1998p133,Friesen:2003p121301,Angus:2007p845,Simmons:2007p213103,Xiao:2010p1876,Morello:2010p687,Simmons:2011p156804} 
because spins in silicon
have been predicted~\cite{Tahan:2002p035314,DeSousa:2003p1358}
and measured~\cite{Tyryshkin:2011unpublished} to have longer coherence times than
spins in many other semiconductors, 
because of both the weak spin-orbit interaction and 
the low nuclear spin density in silicon.
Here, we measure a triplet-singlet relaxation time in a single silicon dot
 to be $>$ 100 ms and demonstrate readout of the singlet and triplet states of
 two electrons in a 
 silicon dot.
We estimate dephasing times theoretically to be on the order of microseconds,
long enough for achieving high fidelity quantum operations.

{\it Qubit design}.   An important advantage of the qubit  proposed here
is that all qubit manipulations can be implemented using electric and not
magnetic fields, resulting in fast operations~\cite{DiVincenzo:2000p1642}.
To understand why electrical manipulation of our
qubit is possible, we
enumerate
the possible transitions between spin states of three electrons that can be induced
by spin-conserving manipulations.
When three spin-1/2 entities are added, 
the resulting 8 total spin eigenstates form a quadruplet with $S = 3/2$ and 
$S_z = 3/2,$ $1/2,$ $-1/2,$ $-3/2,$ and two doublets, each with $S = 1/2,$
$S_z = \pm 1/2$,
where the total spin is $\hbar^2 S(S+1)$ and the z-component of the total spin
is $\hbar S_z$.
Only states with the same $S$ and $S_z$ can be coupled by spin-independent
terms in the Hamiltonian.
We choose to use the group of two states with $S=1/2$, $S_z=-1/2$ for
the states of the qubit.

\begin{figure}
\includegraphics[width=6cm]{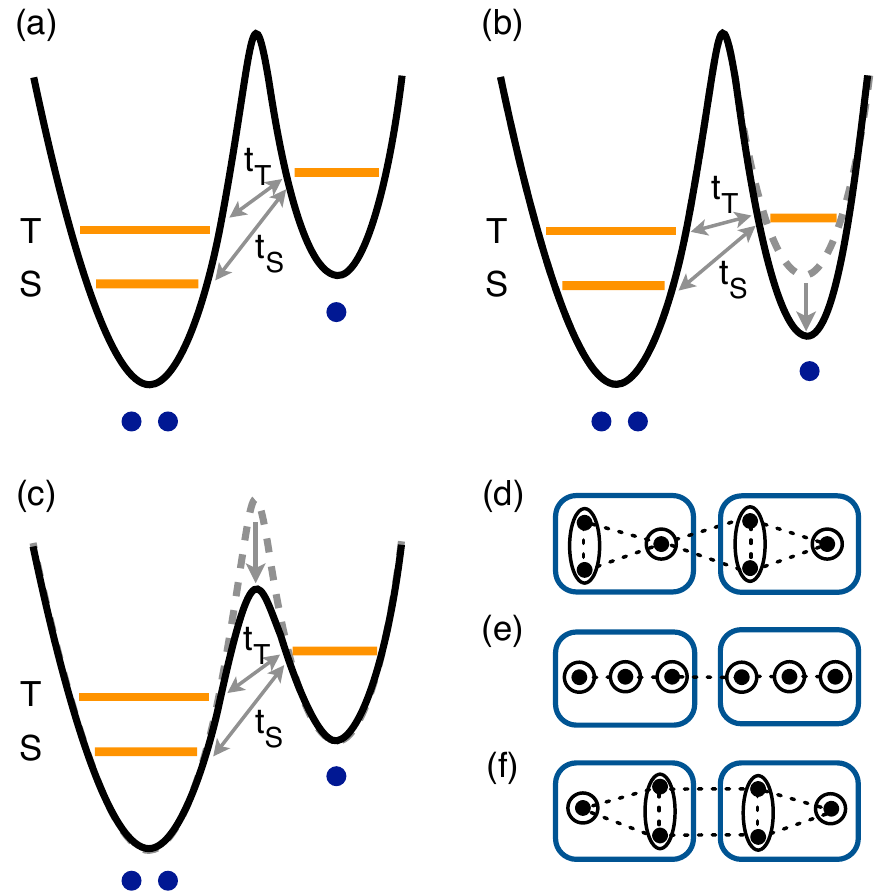}
\caption{\label{fig:schematic}(a) Schematic drawing of a hybrid qubit consisting of three electrons
in two dots.  
The logical qubit states $\ket{0}_L$ and $\ket{1}_L$ both have
$S=1/2,~S_z=-1/2$; they
are $\ket{0}_L=\ket{S}\ket{\downarrow}$ and
$\ket{1}_L=\color{black}{\sqrt{\frac{1}{3}}\ket{T_0} \ket{\downarrow}
- \sqrt{\frac{2}{3}}\ket{T_{-}}\ket{\uparrow}}$, where
$\ket{S}$,
$\ket{T_-}$, and 
$\ket{T_0} $
are two-particle singlet (S) and triplet (T) states in the left dot, and
 $\ket{\uparrow}$ and $\ket{\downarrow}$ respectively
denote a spin-up  and spin-down
 electron in the right dot.
Gate operations are all performed electrically;
gate voltages are used to change the energy splittings between
the singlet and triplet states in the left dot and
to change the tunnel couplings $t_S$ and $t_T$
between the two dots.
(b) and (c): Schematic illustrating that the coupling between
the electron in the singly occupied dot
and the singlet and triplet states in the doubly occupied dot
can be tuned independently via the barrier height and
relative energies in the two dots, as described in the text.
(d): Effective connectivity
of two hybrid qubits composed of four dots
in a linear geometry.  Each connection is a tunable two-electron interaction.  
There are eight effective connections, compared to five effective connections
in a linear array of six dots for the qubits considered in 
Ref.~\cite{DiVincenzo:2000p1642}, shown in (e).
For (d), a two-qubit gate equivalent to CNOT up to local (one-qubit) unitary operations
can be implemented in 16 steps, compared to 18 for (e)~\cite{Fong:2011p1003}
(see Supplemental Information).
(f): Connectivity for which a fourteen-operation two-qubit gate equivalent
to CNOT up to local unitary operations has been
found (see Supplemental Information).
}
\vspace{-.2cm}
\end{figure}

As discussed in~\cite{DiVincenzo:2000p1642}, the two states of the
logical qubit
with $S=1/2$ and $S_z=-1/2$
can be
written as $\ket{0}_L=\ket{S}\ket{\downarrow}$ and
$\ket{1}_L=\color{black}{\sqrt{\frac{1}{3}}\ket{T_0}\ket{\downarrow}
- \sqrt{\frac{2}{3}}\ket{T_{-}}\ket{\uparrow}}$.
In our case,
$\ket{S}$, $\ket{T_0}$, and $\ket{T_-}$ are the singlet
 ($(\ket{\uparrow\downarrow}-\ket{\downarrow\uparrow})/\sqrt{2}$),
$T_0$ triplet ($(\ket{\uparrow\downarrow}+\ket{\downarrow\uparrow})/\sqrt{2}$),
 and $T_-$ triplet ($\ket{\downarrow\downarrow}$) in the left dot, and
 $\ket{\uparrow}$ and $\ket{\downarrow}$ respectively
denote a spin-up  and spin-down
 electron in the right dot.
The essential difference between our system and that of~\cite{DiVincenzo:2000p1642} is that the singlet
and triplet states are of two electrons in one dot instead of two different dots,
as depicted in Fig.~\ref{fig:schematic}.

{\it Single qubit gate operations}.  We now
discuss how gate operations are implemented in this qubit by changing gate voltages
in the device.  A complete set of single-qubit manipulations consists of one that
changes the energy splitting between the qubit states and another that
drives transitions between the qubit states.
The energy difference between the two
qubit states is mainly the singlet-triplet splitting
in the doubly occupied dot, and this splitting indeed can be tuned by changing gate voltages
in both GaAs/GaAlAs~\cite{Amasha:2008p1500} 
and in Si/SiGe dots~\cite{Shi:2011unpublished}.
In Si/SiGe systems, changing the voltage on a global top-gate should also
change  the 
singlet-triplet splitting~\cite{Boykin:2004p165325,Saraiva:2009p081305}.

Transitions between the two states of the hybrid qubit can be induced
by changing the off-diagonal terms in the reduced Hamiltonian.  These terms are each proportional to $t_i^2/\Delta E_i$, where $t_i$ is the relevant
tunneling amplitude and $\Delta E_i$ is the energy difference between the
the relevant state
with two electrons on the left dot
and the virtual state in which an electron has tunneled from state $i$ in the left dot
onto the right dot.
Explicit calculations of the 
effective spin Hamiltonian
obtained by a canonical transformation that systematically eliminates
higher energy
states~\cite{Schrieffer:1966p491,Gros:1987p381,MacDonald:1988p9753}
 demonstrate that increasing the tunnel couplings
between the quantum dots indeed drives transitions between the
two states of the qubit (see Supplemental Information).
Therefore, gate modulations
 that change the $t_i$ will induce transitions between the qubit states, and
modulations of the energy difference $\Delta E_i$ will similarly induce transitions
 when the $t_i$ are non-negligible.
 We note that when the singlet-triplet splitting $\Delta_{ST}$ is  nonzero, 
Rabi flops are performed by modulating the off-diagonal terms at the 
angular frequency $\Omega$ satisfying $\hbar\Omega = \Delta_{ST}$.
This modulation is easier to achieve experimentally when $\Delta_{ST}$ is not
too large.
A singlet-triplet splitting of $0.05$ meV, typical of splittings measured in quantum dots
fabricated in Si/SiGe heterostructures~\cite{Goswami:2007p41,Borselli:2011p123118}, 
corresponds to a frequency of $\sim 10$~GHz.
Quantum dot gate operations have already been achieved at this speed~\cite{Petta:2004p1586}, and efficient schemes exist for refocusing the fast rotations~\cite{Jones:1999}.

While the two manipulations obtained by changing the singlet-triplet splitting in one dot or the tunnel coupling between two dots
described above are sufficient for
achieving arbitrary single qubit gates, a larger set of elementary operations (or, equivalently, more fine-grained control of the terms in the effective Hamiltonian) is useful because
it enables two-qubit gates to be
implemented with fewer elementary
operations.
We note that $t_S$ and $t_T$, the tunneling matrix
elements that shift a single electron from the singlet or triplet state in the left dot to the lowest energy state in the right dot, can be tuned separately. 
Decreasing the tunnel barrier, as shown in Fig.~\ref{fig:schematic}(c),
increases both
tunnel rates, whereas changing the difference between the overall energies
in the
left and right dots, as in Fig.~\ref{fig:schematic}(b), can change the ratio of the two tunnel rates, because
of energy-dependent tunneling~\cite{MacLean:2007p1499,Simmons:2010p245312}.
The tunable degrees of freedom (the
singlet-triplet splitting and
the tunnel rates into the singlet and into the triplet) are denoted schematically in
Fig.~\ref{fig:schematic}(d) as dashed lines.

{\it Two-qubit gates}.  The spin symmetries of the hybrid
qubit are the same as in the three-dot qubit of~\cite{DiVincenzo:2000p1642}
and two-qubit gates are implemented similarly; however, because the hybrid
qubit has higher effective  connectivity,
two-qubit gates can be implemented with fewer elementary operations.
The increased connectivity for dots in a linear array
is illustrated schematically in Fig.~\ref{fig:schematic}(d-f).
Fig.~\ref{fig:schematic}(d) shows two hybrid qubits with eight effective connections, while
Fig.~\ref{fig:schematic}(e) shows the five effective connections of
two triple-dots in a linear array.
Fig.~\ref{fig:schematic}(f) shows a different arrangement of two double-dots, also with
eight effective connections.
We have found sequences of 16 and 14 two-qubit
operations that yield gates equivalent to CNOT up to local unitary
operations for Fig.~\ref{fig:schematic}(d) and Fig.~\ref{fig:schematic}(f), 
respectively.  In comparison, 18 operations are needed
for Fig.~\ref{fig:schematic}(e)~\cite{Fong:2011p1003}.
These shorter gate sequences, presented in the Supplementary Information,
 provide strong evidence
that increased effective connectivity can enable
implementation of gates of two logical qubits
with fewer elementary two-qubit operations.

\begin{figure}
\includegraphics[width=8.5cm]{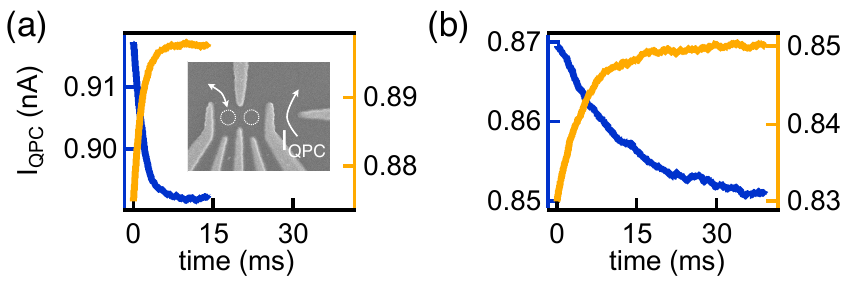}
\caption{\label{fig:tunnel_rates}
Experimental measurement of loading (blue) vs.\ unloading (gold) rates of a Si/SiGe quantum dot in magnetic fields such that the ground state is a triplet (a) or singlet (b).
The current \sym{I}{QPC} through the quantum point contact depends on
the dot occupation, so changes in the current signal loading or unloading.
Following~\protect{\cite{Hanson:2005p719}}, the average occupation is measured as a function of wait time in a pulsed gate experiment.  
The loading and unloading rates for the triplet (singlet) are \amount{521}{Hz} and \amount{645}{Hz} (\amount{81}{Hz} and \amount{182}{Hz}), respectively.  The strong dependence of the tunnel rate on the dot state, triplet or singlet, enables qubit initialization and readout.  Details of the measurement are presented in the supplemental material.
Inset: scanning electron micrograph of a top-gated Si/SiGe dot with the same gate structure as the
one used in the experiment, which is described in \cite{Simmons:2011p156804}.  
}
\end{figure}

{\it Initialization and readout.}  The simplest initialization procedure is to operate
the qubit with a gate voltage tuning and magnetic field chosen so that
 the state $\ket{S}\ket{\downarrow}$ is the lowest energy state of the system, enabling initialization by cooling to the ground state.  Both
initialization and readout can be performed by exploiting different tunnel
 rates in and out of the singlet and triplet states of the doubly occupied dot~\cite{Elzerman:2004p731}.
 Fig.~\ref{fig:tunnel_rates} shows that the tunnel rates of singlets and triplets
 into and out of a Si/SiGe dot differ significantly, and hence
 tunnel rate readout and initialization are feasible in the Si/SiGe system.
 
\begin{figure}
\includegraphics[width=8cm]{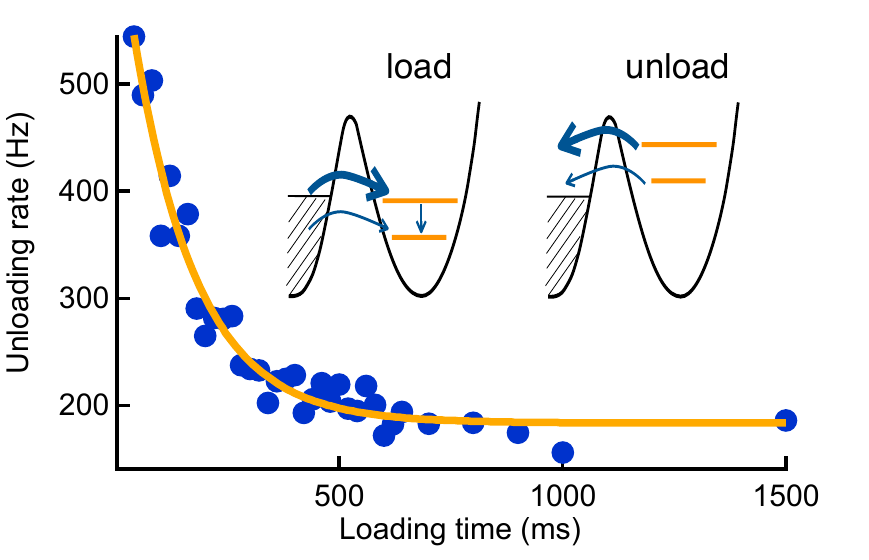}
\caption{\label{fig:relax_rate}Experimental 
measurement of the triplet $\ket{T_-}$ to singlet $\ket{S}$ relaxation time
in the Si/SiGe quantum dot
shown in Fig.~\protect{\ref{fig:tunnel_rates}},
following~\protect{\cite{Hanson:2005p719}}.
The gate voltages on the dot are changed to allow the triplet state to load, and
the unloading rate is measured as a function of the duration of the
loading pulse.
The triplet-to-singlet relaxation time \sym{T}{1}, obtained by fitting the change
in the observed unloading rate to an exponential form (shown as the
solid line), is \amount{141 \pm 12}{ms}.}
\end{figure}

{\it Coherence properties}.  While the spin symmetries of the hybrid qubit
are identical to those of the three-dot qubit in~\cite{DiVincenzo:2000p1642},
the coherence properties are different because the singlet and triplet states
have different spatial wavefunctions.
An essential component of this qubit is a long lifetime for the triplet state of the dot with two electrons.
Fig.~\ref{fig:relax_rate} presents an experimental measurement of the
longitudinal relaxation time $T_1$  of the triplet state $T_-$
in a Si/SiGe dot, yielding $T_1\sim 140$~ms.
This slow relaxation time is significantly longer than the value of
$\sim 3$~ms measured in GaAs~\cite{Hanson:2005p719}
and is consistent
with theoretical estimates that take into account Rashba
spin-orbit coupling and phonon-assisted hyperfine 
coupling~\cite{Prada:2008p1187,Wang:2011p043716}.

The different charge distributions of the two qubit states gives rise to dephasing
due to electron-phonon
 coupling~\cite{Hu:2005p2658,Hu:2011p165322} and
charge noise~\cite{Culcer:2009p073102}.
Our calculations indicate that for realistic states, the intervalley component of the electron-phonon dephasing term
is the most important, and leads to $T_2 \sim 1 $ $\mu$s \cite{Gamble:2011unpub},
yielding $\simgt 10^4$ operations per coherence time.
Dephasing due to charge noise is suppressed in the
hybrid qubit compared to charge qubits~\cite{Hayashi2003p226804}
because the changes in
charge distributions are confined to a single quantum dot, making the effective
dipole moment much smaller (indeed, the dipole moment vanishes in the limit
of harmonic dot potentials)~\cite{Gamble:2011unpub}.
Therefore, charge fluctuation-induced decoherence is greatly suppressed in
the hybrid qubit compared to double dot charge qubits.
Simple estimates indicate that decoherence rates induced by nuclear spins will be similar to
those in singlet-triplet qubits~\cite{Bluhm:2011p109}.

{\it Summary}.  We propose a solid state qubit architecture consisting of three electrons
in two quantum dots.  Compared to previous proposals, this new
qubit has the important advantages of fast gate operations and
relative simplicity of fabrication.  Experimental data
are presented that support the feasibility of constructing
the qubit architecture using Si/SiGe
quantum dots.

We acknowledge useful conversations with Malcolm Carroll, Robert Joynt, and Charles Tahan. This work was supported
in part by ARO and LPS (W911NF-08-1-0482), by NSF
(DMR-0805045, PHY-1104660, and a graduate fellowship to JKG), and by United States Department of Defense.  The US government requires publication of the following disclaimer:  the views and conclusions
contained in this document are those of the authors and should not be
interpreted as representing the official policies, either expressly or
implied, of the US Government. 
This research utilized NSF-supported shared facilities at the University of Wisconsin-Madison.

\begin{onecolumngrid}

\newcounter{subequation}
\renewcommand{\theequation}{S\arabic{subequation}}
\setcounter{subequation}{1}

\newcounter{subfigure}
\renewcommand{\thefigure}{S\arabic{subfigure}}
\setcounter{subfigure}{1}

\newcounter{subtable}
\renewcommand{\thetable}{S\arabic{subtable}}
\setcounter{subtable}{1}

\section*{Supplemental Information}

This Supplementary Information presents details for three topics
 in the main text:
(1) the calculation, performed using a canonical transformation technique, that
shows that transitions between the qubit
states can be driven by modulating the tunnel couplings, which can be
done by application of appropriate gate voltages;
(2) the presentation of gates of two logical qubits that are equivalent to
CNOT up to local unitary operations using gate sequences of 14 and 16 operations 
for two hybrid qubits implemented in four dots in a linear array.
These sequences have fewer operations than
for two three-spin qubits implemented in six dots in a linear array,
for which the shortest sequence known requires 18 operations; and
(3) the presentation of the details of the experiments performed to measure the
lifetime of the $T_-$ triplet state in a Si/SiGe quantum dot.

\section{Derivation of effective Hamiltonian for hybrid qubit using a canonical transformation.}

This section presents the derivation of an effective Hamiltonian that describes the quantum behavior of  the ``hybrid'' double quantum dot spin qubit in the regime of low energy excitations. The ``hybrid'' spin qubit consists of three electrons in a double quantum dot in the (2,1) charge configuration. Changing gate voltages can detune, or change the relative energies of the two dots (while remaining in the regime in which the (2,1) charge configuration has the lowest energy), and also change the tunnel barrier between the dots. We will assume that the gate voltages applied to the quantum dots are kept in the regime in which the the (2,1) charge configuration has lowest energy.   Fig.~\ref{fig:ddot} shows the detuned double quantum dot system with the lowest single particle energies of each dot labeled by 1 and 2.

\begin{figure}
\includegraphics[trim=100 180 100 200,scale=0.6,clip=true]{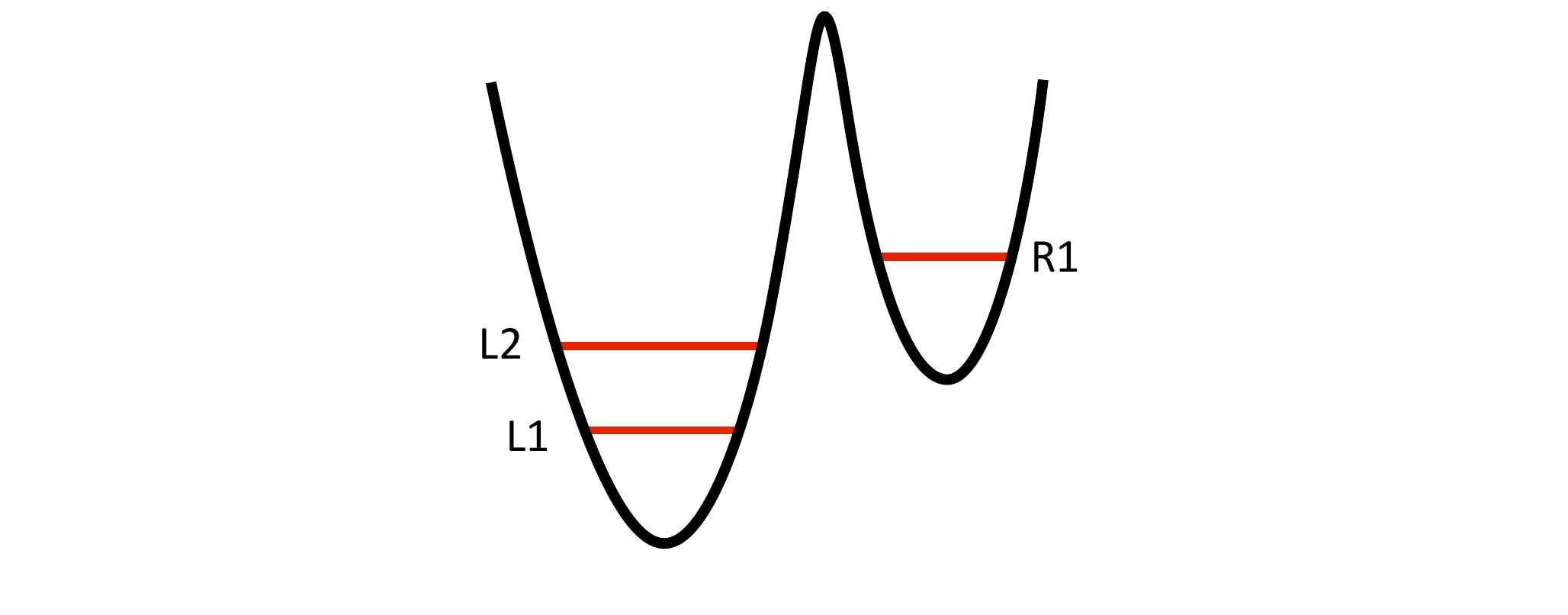}
\vspace{-0.5cm}
\caption{\label{fig:ddot} Schematic of the ``hybrid'' spin qubit. Here, levels L1 and L2 refer to the lowest single particle energies of the left, doubly occupied dot and R1 is the lowest single particle energy of the right, singly occupied dot. We assume that changing appropriate
gate voltages changes the detuning, or relative energies of the states in the two dots (though always with the (2,1) charge configurations having the lowest energy), and that gates can be used to change the tunnel barrier between the quantum dots.
}
\end{figure}
\addtocounter{subfigure}{1}  

Before presenting the details of our calculations, we present in Fig.~\ref{fig:cartoon}
a  cartoon that illustrates why introducing tunneling between dots induces transitions
between the singlet and triplet states of the doubly occupied dot.
Fig.~\ref{fig:cartoon} shows that when tunneling is allowed, there is
a matrix element between the (2,1) state with the two electrons in the left
dot in a singlet and the virtual (1,2) state in which the two electrons in the right
dot are in a singlet, which in turn is coupled to the
(2,1) state with the two electrons in a triplet state.
The coupling between
the (2,1) singlet and (2,1) triplet states is of order $t^2/\Delta$, where $t$ is the tunnel
coupling and $\Delta$ is the energy difference between the (2,1) state and
the (1,2) intermediate state.

\begin{figure}
\includegraphics[trim=0 220 0 180,scale=0.6,clip=true]{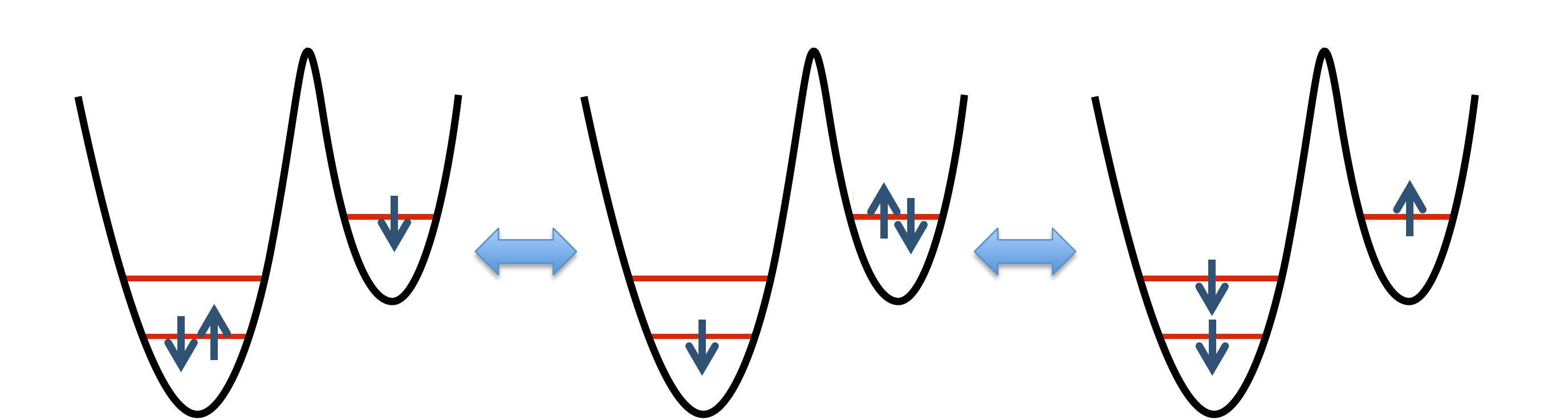}
\caption{\label{fig:cartoon} Cartoon illustrating why introducing tunneling between the
dots induces transitions between the singlet and triplet states in the doubly occupied dot.
Starting from a state in which the electrons are in a singlet, if electron tunnels from the left dot to the right dot, and then the other electron tunnels back to the left dot, the
spins in the left dot will end up in a triplet.  The actual process, which conserves the total $S^2$
and $S_z$, yields the state
$\ket{1}_L=\color{black}{\sqrt{\frac{1}{3}}\ket{T_0}\ket{\downarrow}
- \sqrt{\frac{2}{3}}\ket{T_{-}}\ket{\uparrow}}$.
}
\end{figure}
\addtocounter{subfigure}{1}  

We now provide a more detailed calculation of the couplings between
the states of the hybrid qubit.
The Hamiltonian describing interacting electrons, in general, can be written in the Hubbard-like form
\begin{equation}\label{eq:H}
\hat{H} =  \sum_{\alpha,i,s} {\left( \epsilon_{\alpha i} + \mu_\alpha \right) \ \hat{n}_{\alpha i s}  } 
+ \sum_{\alpha \neq \beta} \sum_{i,j} \sum_{s} {t_{\alpha i, \beta j} \ \hat{c}^\dag_{\alpha i s} \hat{c}_{\beta j s}} 
+  \frac{1}{2}\sum_{\alpha, \beta, \gamma, \delta}  \sum_{i,j,k,l} \ \sum_{s,s'} { \ \Gamma^{\alpha \beta \gamma \delta}_{i j k l} \ \hat{c}^\dag_{\alpha i s} \hat{c}^\dag_{\beta j s'} \hat{c}_{\gamma k s'} \hat{c}_{\delta l s}  },
\end{equation}
\addtocounter{subequation}{1}
where we have used Greek indices $\alpha, \beta, \gamma, \delta$ to refer to the quantum dots, Roman indices $i, j, k, l$  to refer to the orbitals of a quantum dot, and $s, s'$ are spin indices. In this notation, operator $\hat{c}^\dag_{\alpha i s}$ ($\hat{c}_{\alpha i s}$) creates (annihilates) an electron with spin $s$ in the $i$-th orbital of the $\alpha$-th quantum dot, and $\hat{n}_{\alpha i s} = \hat{c}^\dag_{\alpha i s} \hat{c}_{\alpha i s}$ is the number operator. 

In the Hamiltonian of Eq.~\ref{eq:H}, the single particle energy of the $i^{th}$ orbital in the $\alpha^{th}$ quantum dot is given by $\epsilon_{\alpha i}$, and the energy shift of the quantum dot is given by $\mu_\alpha$. This energy shift can be controlled experimentally by application of some gate voltage. 
The second term on the right-hand side of Eq.~\ref{eq:H} is the tunneling operator, which describes the tunneling of an electron with spin $s$ between the $j$-th orbital of the $\beta^{th}$ quantum dot and the $i^{th}$ orbital of the $\alpha^{th}$ quantum dot, where $\alpha$ and $\beta$ refer to different quantum dots. 
Orbital and tunneling energies can be calculated by single particle integrals between quantum dot orbital wavefunctions $\phi_{\alpha i}(\mathbf{r})$,
\begin{eqnarray}
\epsilon_{\alpha i} &=& \int d\mathbf{r} \ \phi^*_{\alpha i }(\mathbf{r}) \left[ \frac{1}{2m^*}\mathbf{\hat{p}}^2 + V_p(\mathbf{r})  \right] \phi_{\alpha i } (\mathbf{r}),\\
\addtocounter{subequation}{1}
t_{\alpha i, \beta j}  &=& \int d\mathbf{r} \ \phi^*_{\alpha i }(\mathbf{r}) \left[ \frac{1}{2m^*}\mathbf{\hat{p}}^2 + V_p(\mathbf{r})  \right] \phi_{\beta j } (\mathbf{r}),
\end{eqnarray}
\addtocounter{subequation}{1}
where $V_p(\mathbf{r})$ is the model quantum dot confinement potential, $m^*$ is the effective mass of an electron in the conduction band, $e$ is the electronic charge and $\hat{\mathbf{p}}$ is the momentum operator.

The prefactor of the third term in the Hamiltonian in Eq.~\ref{eq:H} is a two-electron Coulomb integral between quantum dot orbital wavefunctions $\phi(\mathbf{r})$, given by
\begin{equation}\label{eq:Gamma}
\Gamma^{\alpha \beta \gamma \delta}_{i j k l} = \iint \ d\mathbf{r} \ d\mathbf{r}' \ 
\phi^*_{\alpha i } (\mathbf{r})  \phi^*_{\beta j } (\mathbf{r}') \frac{e^2}{4\pi \epsilon_r \epsilon_0 |\mathbf{r} - \mathbf{r}' |} \phi_{\gamma k} (\mathbf{r}')  \phi_{\delta l} (\mathbf{r}),
\end{equation}
\addtocounter{subequation}{1}
where $\epsilon_0$ is the permittivity of free space and $\epsilon_r$ the relative permittivity of the material.

When the quantum dot indices refer to the same dot ($\alpha=\beta=\gamma=\delta$), we get the intra-dot Coulomb energies. This can be further categorized into the direct and exchange Coulomb energies. When the orbital indices $i=l$ and $j=k$, we obtain the direct Coulomb energy, $\Gamma^{\alpha\alpha\alpha\alpha}_{ijji} \equiv C_{\alpha i,\alpha j}$. When the orbital indices $i=k$ and $j=l$, we get the exchange Coulomb energy, $\Gamma^{\alpha\alpha\alpha\alpha}_{ijij} \equiv K_{\alpha i,\alpha j}$. When the quantum dot indices refer to the different dots ($\alpha=\delta, \ \beta=\gamma, \ \alpha \neq \beta$), we obtain the direct and exchange Coulomb energies between the electrons localized in the different quantum dots, i.e. $\Gamma^{\alpha\beta\beta\alpha}_{ijji} \equiv C_{\alpha i,\beta j}$ and $\Gamma^{\alpha\beta\beta \alpha}_{ijij} \equiv K_{\alpha i,\beta j}$.

These direct and exchange Coulomb energies are given by
\begin{eqnarray}
C_{\alpha i, \beta j} &=& \iint \ d\mathbf{r} \ d\mathbf{r}' \ 
\phi^*_{\alpha i } (\mathbf{r})  \phi^*_{\beta j } (\mathbf{r}') \frac{e^2}{4\pi \epsilon_r \epsilon_0 |\mathbf{r} - \mathbf{r}' |} 
\phi_{\beta j} (\mathbf{r}') \phi_{\alpha i} (\mathbf{r})  , \\
\addtocounter{subequation}{1}
K_{\alpha i, \beta j} &=& \iint \ d\mathbf{r} \ d\mathbf{r}' \ 
\phi^*_{\alpha i } (\mathbf{r})  \phi^*_{\beta j } (\mathbf{r}') \frac{e^2}{4\pi \epsilon_r \epsilon_0 |\mathbf{r} - \mathbf{r}' |}
 \phi_{\beta i} (\mathbf{r}')  \phi_{\alpha j} (\mathbf{r}) .
\end{eqnarray}
\addtocounter{subequation}{1}

Because of the non-vanishing overlap of the orbital wavefunctions between different quantum dots in the regime where the dots are coupled, other integrals also yield nonzero contributions. In general, for a system of two quantum dots, the integrals $\Gamma^{\alpha\beta\beta\alpha}_{ijkl}$ and $\Gamma^{\beta\alpha\beta\alpha}_{ijkl}$ may be non-zero for different orbitals $i,j,k$ and $l$, due to the overlap between orbitals centered on different quantum dots $\alpha$ and $\beta$. On the other hand, for a single quantum dot, if we assume a symmetric confinement potential $V_p({\bf{r}})$, such terms vanish.

By categorizing these energies into intra-dot, inter-dot and tunneling energies, we can transform the Hamiltonian of Eq.~\ref{eq:H} into a more familiar form~\cite{DasSarma:2011p235314} with intra-dot energy, inter-dot energy and tunneling operators $\hat{U}_0$, $\hat{U}_1$ and $\hat{T}$, respectively:
\begin{equation}\label{eq:Hubbard}
\hat{H} = \hat{U}_0 + \hat{U}_1 + \hat{T},
\end{equation}
\addtocounter{subequation}{1}
where
\begin{eqnarray}
\hat{U}_0 &=&   \sum_{\alpha,i,s} {\left( \epsilon_{\alpha i} + \mu_\alpha \right) \ \hat{n}_{\alpha i s}  }  +   \frac{1}{2}\sum_{\alpha}  \sum_{i,j} \ \sum_{s,s'} {\left( \ C_{\alpha i,\alpha j} \ \hat{c}^\dag_{\alpha i s} \hat{c}^\dag_{\alpha j s'} \hat{c}_{\alpha j s'} \hat{c}_{\alpha i s} 
+ K_{\alpha i,\alpha j} \ \hat{c}^\dag_{\alpha i s} \hat{c}^\dag_{\alpha j s'} \hat{c}_{\alpha i s'} \hat{c}_{\alpha j s} \right)}, \label{eq:U} \\
\addtocounter{subequation}{1}
\hat{U}_1 &=&  \frac{1}{2}\sum_{\alpha \neq \beta} \sum_{ \gamma \neq \delta}  \sum_{i,j,k,l} \ \sum_{s,s'} { \ \Gamma^{\alpha \beta \gamma \delta}_{i j k l} \ \hat{c}^\dag_{\alpha i s} \hat{c}^\dag_{\beta j s'} \hat{c}_{\gamma k s'} \hat{c}_{\delta l s}  },\\
\addtocounter{subequation}{1}
\hat{T} &=& \sum_{\alpha \neq \beta} \sum_{i,j} \sum_{s} {t_{\alpha i, \beta j} \ \hat{c}^\dag_{\alpha i s} \hat{c}_{\beta j s}} \label{eq:T} .
\end{eqnarray}
\addtocounter{subequation}{1}

The Hilbert space of a system of three electrons confined in a double quantum dot potential spans the space of states with charge configurations (3,0), (0,3), (2,1) and (1,2). Here, number pairs $(n, m)$ denote the number of electrons in the left $(n)$ and right $(m)$ dots.
We shall denote by $\mathcal{A}^{(2,1)}$, $\mathcal{A}^{(1,2)}$, $\mathcal{A}^{(3,0)}$ and $\mathcal{A}^{(0,3)}$ the set of states spanning the subspaces of (2,1), (1,2), (3,0) and (0,3) states respectively. The Hilbert space of the ``hybrid'' qubit is given by direct sum of these subspaces, and will be denoted by $\mathcal{A} \equiv \mathcal{A}^{(2,1)} \oplus \mathcal{A}^{(1,2)}  \oplus \mathcal{A}^{(3,0)}  \oplus \mathcal{A}^{(0,3)}$.  In the Hilbert space $\mathcal{A}$, the inter-dot tunneling operator $\hat{T}$ describes the coherent tunneling of electrons between states of the distinct subspaces whereas the intra-dot and inter-dot energy operators $\hat{U}_0$ and $\hat{U}_1$ do not couple these subspaces.

We can treat the tunneling matrix elements perturbatively and block-diagonalize the Hamiltonian by using the Schrieffer-Wolff 
transformation~\cite{Schrieffer:1966p491,Gros:1987p381,MacDonald:1988p9753}. In this way we can transform the Hubbard-like Hamiltonian into a spin Hamiltonian $\hat{H}'$ by a unitary transformation,
\begin{eqnarray}
\hat{H}' &=& e^{i \hat{S}} \hat{H} e^{-i \hat{S}} \nonumber \\
   &=& \hat{H} + [ i\hat{S}, \hat{H} ] + \ldots \label{eq:exp} \nonumber \\
   &=& ( \hat{U}_0 + \hat{U}_1) + \hat{T} + [ i\hat{S},  ( \hat{U}_0 + \hat{U}_1)] + [ i\hat{S}, \hat{T} ] + \ldots \label{eq:comm}
\end{eqnarray}
\addtocounter{subequation}{1}

We note that $i \hat{S}$ can be chosen such that $\hat{T} + [ i\hat{S}, ( \hat{U}_0 + \hat{U}_1)] = 0$, yielding an effective Hamiltonian for states with a fixed number of electrons on each dot. The transformed Hamiltonian becomes, in the lowest order, $\hat{H}' \approx \hat{U}_0 + \hat{U}_1 + [i\hat{S}, \hat{T} ] $. Ref.~\onlinecite{Gros:1987p381} provides an explicit expression for the choice of $i \hat{S}$, given by
\begin{equation}
i\hat{S} = \sum_{n,m} | \varphi_n \rangle \frac{\langle \varphi_n |  \hat{T}  |  \varphi_m \rangle}{ \langle \varphi_n |\hat{U} | \varphi_n \rangle - \langle \varphi_m | \hat{U} | \varphi_m \rangle} \langle  \varphi_m |,
\end{equation}
\addtocounter{subequation}{1}
where the states $ | \varphi_n \rangle \in \mathcal{A} $ are eigenstates of $\hat{U} \equiv  \hat{U}_0 + \hat{U}_1$.
Since $| \varphi_n \rangle$ are eigenstates of $\hat{U}$,  from the eigenvalue equation, $\hat{U} | \varphi_n \rangle = U_n | \varphi_n \rangle$, the orthogonality condition $\langle \varphi_n | \varphi_m \rangle = \delta_{nm}$, and the completeness relation, $ \sum_{n} | \varphi_n \rangle \langle \varphi_n | = \hat{\mathbf{1}}$, one obtains
\begin{eqnarray}
[i\hat{S}, \hat{U}] &=&  - \hat{T}.
\end{eqnarray}
\addtocounter{subequation}{1}

The first order commutator  $[ i\hat{S}, \hat{T} ]$ gives
\begin{eqnarray}\label{eq:commutator}
&&\langle \varphi_k | [ i\hat{S}, \hat{T} ] |\varphi_{k'} \rangle \nonumber \\
&& = \sum_{m} \frac{ \langle \varphi_k | \hat{T}  |  \varphi_m \rangle \langle \varphi_m | \hat{T}  |  \varphi_{k'} \rangle }{U_k - U_m} - \sum_{ m} \frac{ \langle \varphi_k | \hat{T}  |  \varphi_m \rangle \langle \varphi_m | \hat{T}  |  \varphi_{k'} \rangle }{U_m - U_{k'}}~.  \label{eq:hop2}
\end{eqnarray}
\addtocounter{subequation}{1}

Therefore, the first order commutator contains terms describing the second order hopping of an electron between states in the subspaces. These second order processes produce hoppings between the subspaces as follows: $\mathcal{A}^{(0,3)} \rightarrow \mathcal{A}^{(1,2)} \rightarrow \mathcal{A}^{(0,3)}, \mathcal{A}^{(3,0)} \rightarrow \mathcal{A}^{(2,1)} \rightarrow \mathcal{A}^{(3,0)}, \mathcal{A}^{(1,2)} \rightarrow \mathcal{A}^{(0,3)} \rightarrow \mathcal{A}^{(1,2)},  \mathcal{A}^{(2,1)} \rightarrow \mathcal{A}^{(3,0)} \rightarrow \mathcal{A}^{(2,1)}, \mathcal{A}^{(1,2)} \rightarrow \mathcal{A}^{(2,1)} \rightarrow \mathcal{A}^{(1,2)}$ and $ \mathcal{A}^{(2,1)} \rightarrow \mathcal{A}^{(1,2)} \rightarrow \mathcal{A}^{(2,1)}.$

Eq.~\ref{eq:hop2} shows that second order processes have coefficients that are are inversely proportional to the magnitude of the intra-dot energy difference $ | U_m - U_k |$ between states $| \varphi_m \rangle$ and $| \varphi_k \rangle$.  Since intra-dot energies of the states in subspaces $\mathcal{A}^{(0,3)}$ and $\mathcal{A}^{(3,0)}$ are larger than those of the states in subspaces $\mathcal{A}^{(2,1)}$ and $\mathcal{A}^{(1,2)}$ because of the much larger Coulomb repulsion between electrons in triply-occupied quantum dots, second order hopping processes involving triply-occupied states are virtual processes with much smaller probability compared to hoppings between states in the subspaces $\mathcal{A}^{(2,1)}$ and $\mathcal{A}^{(1,2)}$. As we are interested in the low-lying energy states of the system, we can restrict our basis to the states in the subspace of $\mathcal{A}^{(2,1)} \oplus \mathcal{A}^{(1,2)}$ and ignore triply-occupied states.

Because the qubit states are encoded in the $S=\frac{1}{2}, S_z = -\frac{1}{2}$ two-level subspace belonging to $\mathcal{A}^{(2,1)}$, we can preserve the (2,1) charge configuration of the qubit by adjusting the gate voltages that define the dots so that the lowest energy $\mathcal{A}^{(1,2)}$ configuration has  higher energy than both of the qubit states in $\mathcal{A}^{(2,1)}$. 
The energy coefficient of the $n$-th order of the commutators in Eq.~\ref{eq:comm} is proportional to $\frac{t^{n+1}}{(U_m - U_k)^n}$ and the series in Eq.~\ref{eq:comm} can be truncated to desired order when $t \ll \ | U_m - U_k|$. Here, we keep only the lowest order contribution.

The effective Hamiltonian that describes the low-lying energy states can then be projected onto the subspace $\mathcal{A}^{(2,1)}$ by the projection operator $\hat{P}^{(2,1)} \equiv \sum_{k}  |\varphi_k^{(2,1)} \rangle \langle \varphi_k^{(2,1)}  |$, where the set of states $\{ | \varphi_k^{(2,1)}  \rangle \}$ span $\mathcal{A}^{(2,1)}$. The effective Hamiltonian is given by
\begin{eqnarray}
\hat{H}_{\text{eff}} &=& \hat{P}^{(2,1)} \hat{H}' \hat{P}^{(2,1)} \nonumber \\
&= & \hat{P}^{(2,1)} \left(  \hat{U} + [i\hat{S}, \hat{T}] \right) \hat{P}^{(2,1)} ~.
\end{eqnarray}
\addtocounter{subequation}{1}

With the understanding that we are restricting our basis to the $\mathcal{A}^{(2,1)}$ subspace, we drop the projection operators $\hat{P}^{(2,1)}$.
Using the bilinear spin identity $\vec{\mathbf{S}}_i = \frac{1}{2} \sum_{s,s' }\hat{c}^\dag_{is} \vec{\sigma}_{ss'} \hat{c}_{is'}$, we cast the effective Hamiltonian into the form
\begin{equation}\label{eq:tJ}
\hat{H}_\text{eff}^{\text{(t-J)}} = \hat{U} + \hat{H}_J + \hat{H}_{J'},
\end{equation}
\addtocounter{subequation}{1}
reminiscent of the t-J model~\cite{Auerbach:1998}. The terms $\hat{H}_J$ and $\hat{H}_{J'}$ are  in the form of inter-dot and intra-dot Heisenberg-like exchange interactions, and are given explicitly by
\begin{eqnarray}\label{eq:H_J}
\hat{H}_{J}&=& \frac{1}{2}\sum_{i}J_{i}\left[\left(\sum_{s,s'} \hat{c}^{\dag}_{L i s}\vec{\sigma}_{s s'} \hat{c}_{L i s'} \right) \cdot\left(\sum_{s,s'} \hat{c}^{\dag}_{R1s}\vec{\sigma}_{ss'} \hat{c}_{R1s'} \right) -\hat{n}_{Li} \hat{n}_{R1}\right] \nonumber \\
&=& \frac{1}{2}\sum_{i}J_{i}(4\vec{\mathbf{S}}_{L i}\cdot\vec{\mathbf{S}}_{R1} -\hat{n}_{Li} \hat{n}_{R1}),\\
\addtocounter{subequation}{1}
\hat{H}_{J'}&=&\frac{1}{2}\sum_{i \neq j}J'_{i j}\left[\left(\sum_{s,s'} \hat{c}^{\dag}_{L i s}\vec{\sigma}_{s s'} \hat{c}_{Lj s'} \right) \cdot\left(\sum_{s,s'} \hat{c}^{\dag}_{R1s}\vec{\sigma}_{ss'} \hat{c}_{R1s'} \right) -\left(\sum_{s} \hat{c}^{\dag}_{Li s} \hat{c}_{Ljs} \right) \hat{n}_{R1}\right] \label{eq:H_J'}
\end{eqnarray}
\addtocounter{subequation}{1}
Labels $L, R$ denote the left and right quantum dots and $ i , j = 1,2 $ refer to the lowest two orbitals of a quantum dot. 
The meaning of these operators are as follows. $\hat{H}_J$ is the Heisenberg exchange coupling between electron spin in the $i^{th}$ orbital of the left quantum dot and the electron spin in the ground  orbital of the right quantum dot (R1). $\hat{H}_{J'}$ is the exchange coupling between the electron spins in the $i^{th}$ and $j^{th}$ orbital of the left quantum dot, mediated by the spin in the ground orbital of the right quantum dot.

\begin{figure}
\includegraphics[trim=100 190 100 180,scale=0.6,clip=true]{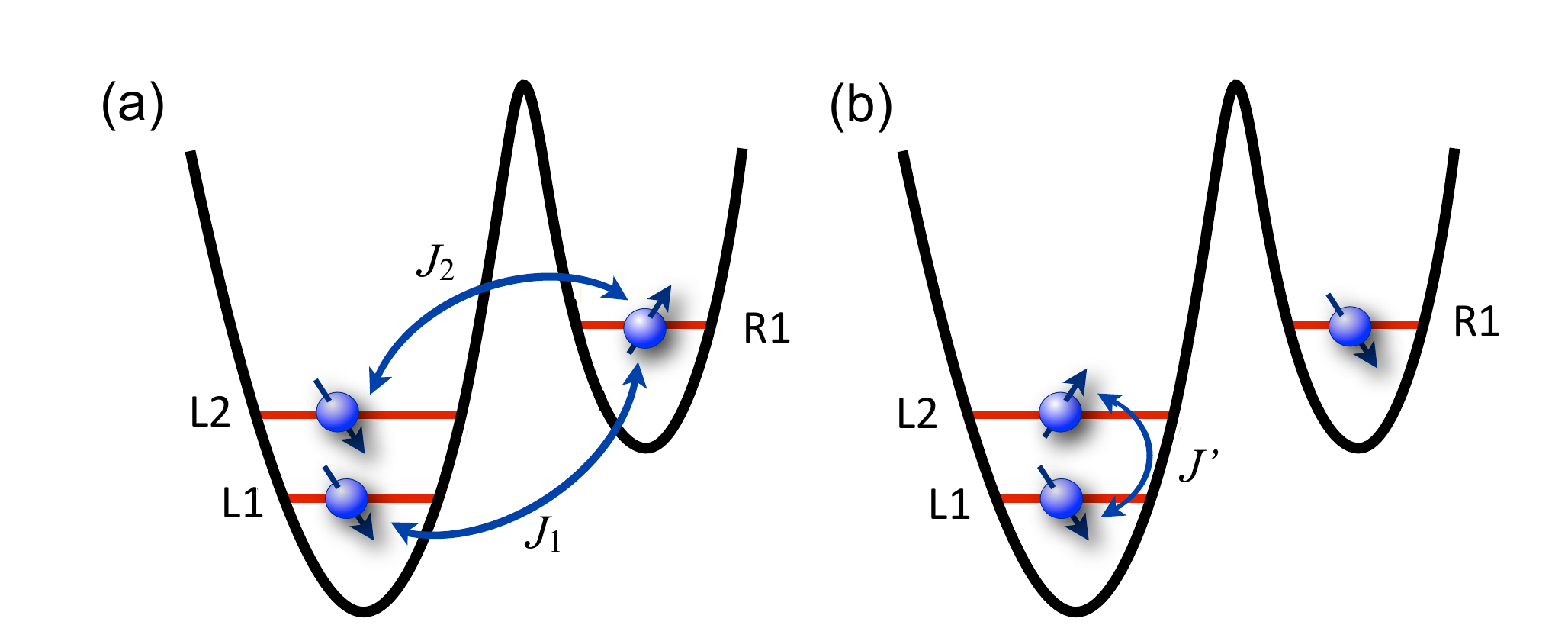}
\caption{\label{fig:J} Schematic of the Heisenberg exchange couplings as given by Eqs.~\ref{eq:H_J} and \ref{eq:H_J'}.  Levels L1 and L2 refer to the lowest single particle energies of the left quantum dot, and R1 refers to the lowest single particle energy of the right dot. (a) $J_1$ and $J_2$ are respectively, the Heisenberg exchange couplings between electron spins in the ground (1) and first excited (2) orbitals of the left quantum dot and the electron spin in the ground  orbital of the right quantum dot. (b) $J'$ is the exchange coupling between the electron spins in the lowest two orbitals of the left quantum dot, mediated by the spin in the ground orbital of the right quantum dot. Explicit forms of $J_1$, $J_2$ and $J'$ are given in Eqs.~\ref{eq:J} assuming equal tunneling energies between the orbitals of the two quantum dots.
}
\end{figure}
\addtocounter{subfigure}{1}

Here, we note that the difference in the energy eigenvalues $U_m$ and $U_k$, of the total intra- and inter-dot energy operator $\hat{U} \equiv \hat{U}_0 + \hat{U}_1$  in the subspaces of $\mathcal{A}^{(2,1)}$ and $\mathcal{A}^{(1,2)}$ appear in the denominator of each of the terms in Eq.~\ref{eq:hop2}. If $m$ and $k$ refer to the ground eigenstates of these subspaces, the biggest contribution to the energy difference in the denominator would be due to the detuning or energy shift differences between the dots, followed contributions from differences in intra-dot energies, and finally, from differences in inter-dot energies in states $m$ and $k$.  Therefore, one could omit the inter-dot energy operator $\hat{U}_1$ from the Hubbard-like Hamiltonian (Eq.~\ref{eq:Hubbard}) from the start, without changing the physical processes described by the t-J Hamiltonian (Eq.~\ref{eq:tJ}) as it contributes only a small correction to the Heisenberg coupling constants $J_i$ and $J'_{ij}$.
In order to make our model less complicated without losing the essential physics, we make the approximation of neglecting these inter-dot contributions. We also assume that tunneling energies for different orbitals in Eq.~\ref{eq:T} are equal, $t_{\alpha i,\beta j} \equiv t$. 

Noting that the ground state of two electrons in a quantum dotat low magnetic fields  is a singlet and the next three higher energy states are triplets and denoting the energies of the singlet (S) and triplet (T) states of the left (L) and right (R) quantum dot by $E_{S/T}^{L/R}$, we can then write the forms of $J_i$ and $J'_{ij}$ in terms of the tunneling energies and singlet and triplet energies. They are given by
\begin{eqnarray}
J_1 \approx \frac{2t^2 }{ E_S^R - E_S^L} ,  && \
J_2 \approx \frac{ 2t^2} {E_S^R - E_T^L }, \nonumber \\
J_{12}'  = J_{21}' &=& \frac{J_1 + J_2}{2} \equiv J'. \label{eq:J}
\end{eqnarray}
\addtocounter{subequation}{1}
The couplings due to these exchange terms are shown schematically in Fig.~\ref{fig:J}. Here, we assume that $E_S^R - E_T^L > 0$ is satisfied  due to detuning, tighter right quantum dot confinement or some experimental scheme that is employed. Then, $E_S^R - E_S^L > 0$ follows because the triplet states are higher in energy than the singlet state, $E_T^L > E_S^L$.
Note that in truncating the series in Eq.~\ref{eq:comm}, we also implicitly assumed that $t <( E_S^R - E_T^L) $.

For the form of intra-dot energy operator given in Eq.~\ref{eq:U}, the singlet and triplet energies are
\begin{eqnarray}
E_S^L &=& 2\epsilon_{L1} + C_{L1,L1} + 2\mu_L, \nonumber \\
E_S^R &=& 2\epsilon_{R1} + C_{R1,R1} + 2\mu_R, \nonumber \\
E_T^L &=& \epsilon_{L1} + \epsilon_{L2} + C_{L1,L2} - K_{L1,L2} + 2\mu_L,\nonumber \\
E_T^R &=& \epsilon_{R1} + \epsilon_{R2} + C_{R1,R2} - K_{R1,R2} + 2\mu_R.
\end{eqnarray}
\addtocounter{subequation}{1}

The logical basis given in the main text is in spin form. To make it consistent with the notations introduced here, we explicitly write them out below. Here, the three-electron kets are the usual Slater determinants.
\begin{eqnarray}
|0\rangle_{\text{L}} & \equiv &|S\rangle |\downarrow\rangle 
 =  | \phi_{L1 \uparrow}(\mathbf{r}_1) \ \phi_{L1\downarrow} (\mathbf{r}_2) \ \phi_{R1\downarrow} (\mathbf{r}_3) \rangle , \\
 \addtocounter{subequation}{1}
|1\rangle_{\text{L}} &\equiv& \sqrt{\frac{1}{3}}|T_0\rangle |\downarrow\rangle - \sqrt{\frac{2}{3}} |T_-\rangle |\uparrow\rangle \nonumber \\  
& = & \sqrt{\frac{1}{6}} \left[ | \phi_{L1\uparrow}(\mathbf{r}_1) \ \phi_{L2\downarrow} (\mathbf{r}_2) \ \phi_{R1\downarrow} (\mathbf{r}_3) \rangle +  | \phi_{L1\downarrow}(\mathbf{r}_1) \ \phi_{L2\uparrow} (\mathbf{r}_2) \ \phi_{R1\downarrow} (\mathbf{r}_3) \rangle \right]
- \sqrt{\frac{2}{3}} | \phi_{L1\downarrow}(\mathbf{r}_1) \ \phi_{L2\downarrow} (\mathbf{r}_2) \ \phi_{R1\uparrow} (\mathbf{r}_3) \rangle  .
\nonumber\\
\end{eqnarray}
\addtocounter{subequation}{1}

The effective Hamiltonian in the logical basis $\{ |0\rangle_{\text{L}}  , |1\rangle_{\text{L}}  \}$ is given by
\begin{equation}\label{effectiveHam}
\hat{H}_{\text{eff}}^{\text{(t-J)}} = \left(
\begin{array}{cc}
 -J_1 & \sqrt{\frac{3}{2}} J'  \\
 \sqrt{\frac{3}{2}} J' & E_{ST}^L -\frac{3}{2}   (J_1 + J_2) 
\end{array}
\right),
\end{equation}
\addtocounter{subequation}{1}
with $E_{ST}^L \equiv E_T^L - E_S^L $ being the singlet-triplet splitting of the two electrons in the left quantum dot. Of significance is the exchange term $J'$ whose expression is given in Eq.~\ref{eq:J}, which couples the two logical states of the qubit and can be modulated by means of electrically controlling the tunnel barrier between the quantum dots.

\section{CNOT gates for qubits with increased connectivities.}
This section presents gate sequences that implement gates 
that are equivalent to CNOT, up to local unitary transformations,
for two logical qubits that consist of the two eigenstates of
three spins with $J=1/2$, $J_z=-1/2$.
As noted in the main text, the spin symmetries of the hybrid
qubit proposed here and the triple-dot qubit of Ref.~\onlinecite{DiVincenzo:2000p1642}
are the same.
However, two hybrid qubits consisting of six electrons in four
dots arranged in a linear array has
 more spin-spin couplings than two logical qubits consisting
of six electrons in six dots arranged in a linear array, as shown in Fig. 1 of the main text.
Here we present evidence that this increased connectivity can
be exploited to reduce the number of gate operations needed to implement
gates of two logical qubits, by presenting two-spin gate sequences that are equivalent
to a CNOT gate, up to local unitary operations, for different connectivities
of the spin-$1/2$ particles that make up the logical qubits.
This result is not unexpected, since  Ref.~\onlinecite{DiVincenzo:2000p1642} shows that for single qubit operations, at most three gates are needed when all three spins are coupled (e.g. in a triangular quantum dot configuration), and at most four gates are necessary when the spins are coupled only to nearest neighbors in a linear array.

The quantum dot geometries we considered are shown in panel (a) of Figs.~\ref{fig:cnot1} and~\ref{fig:cnot2}.
We implemented a search algorithm similar to the one described in Ref.~\onlinecite{Fong:2011p1003}, which is a combination of Nelder-Mead~\cite{Nelder:1965p308} and genetic~\cite{Goldberg:1989} algorithms.  We parallelized the program and exploited the resources of the University of Wisconsin Center for High Throughput Computing (CHTC)~\cite{CHTC} grid using Condor~\cite{Condor} to improve the speed of the calculations. As in Ref.~\onlinecite{Fong:2011p1003}, the numerical results could be used to deduce exact gate times, which points to the power of the algorithm and perhaps some underlying hidden symmetry of the $S, S_z=1$ subspace in which the two logical qubits reside. The exact gate sequences are presented schematically in Fig.~\ref{fig:cnot1}(b) and Fig.~\ref{fig:cnot2}(b). 

We also obtained the local unitary operations necessary for obtaining the exact CNOT gate for the geometry of Fig.~\ref{fig:cnot2}(a), and the exact times and spin couplings are presented in Fig.~\ref{fig:cnot2}(b). 
Note that in this coded qubit scheme, the local unitary operations are
implemented as exchange gates.

\begin{figure}
\includegraphics[scale=0.5]{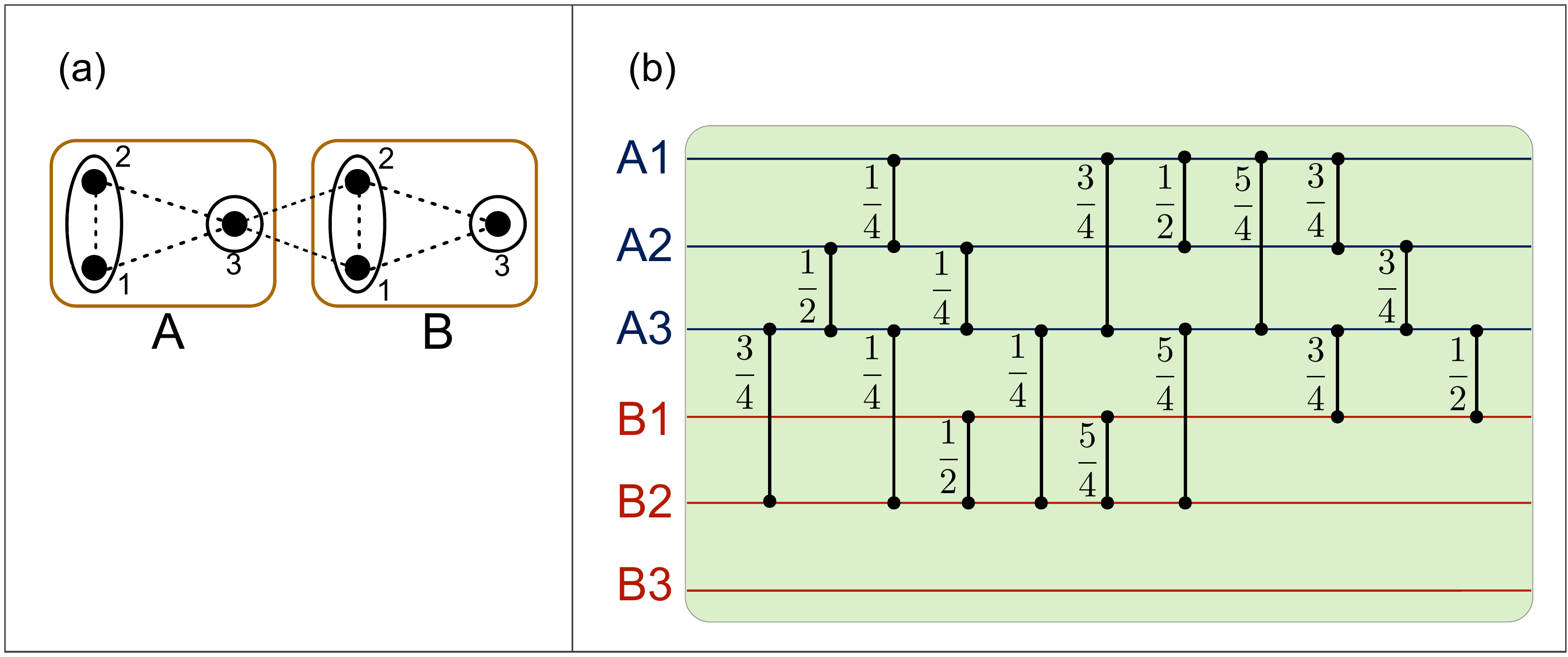}
\caption{\label{fig:cnot1} (a) Schematic of qubit-qubit couplings used to implement a gate that is equivalent to CNOT, up to local unitary operations~\cite{Makhlin:2002p243}. Each logical qubit A and B, is encoded in three spins 1, 2 and 3, with spins 1 and 2 both in one quantum dot.
(b) A 16 gate sequence, with gate times indicated,  that implements a gate that is equivalent to CNOT up to local unitary transformations between logical qubits A and B, in 11 time steps. The order of gate couplings run from left to right. Each coupling between the $i^{th}$ and $j^{th}$ spins represents the unitary evolution $U_{ij}(t) = \mathrm{exp}\left(-\frac{i}{\hbar} t J \vec{\mathbf{S}}_i \cdot \vec{\mathbf{S}}_j  \right)$. Gate times are in units of $h/J$, so that gate times of $\frac{1}{4}$ and $\half$ correspond to $\sqrt{\textrm{SWAP}}$ and SWAP operations respectively. 
}
\end{figure}
\addtocounter{subfigure}{1}

\begin{figure}
\includegraphics[scale=0.5]{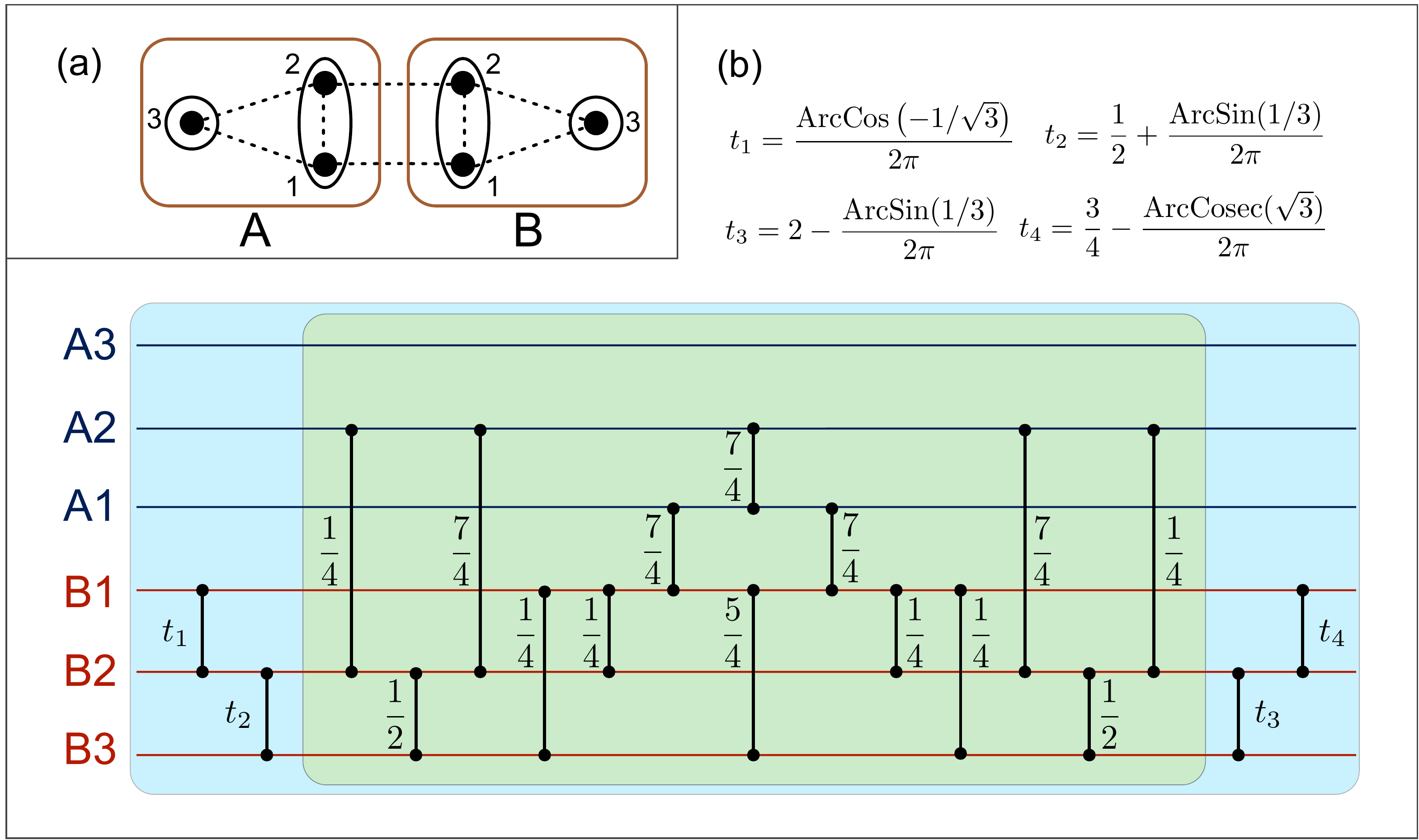}
\caption{\label{fig:cnot2} (a) Schematic of qubit-qubit couplings used to implement a CNOT gate of two logical qubits. Each logical qubit A and B, is encoded in three spins 1,2 and 3, with spins 1 and 2 both in one quantum dot.
(b) Gate sequence, with gate times indicated, that implements an exact CNOT gate between logical qubits A and B, with the order of the couplings running from left to right. The central block (in green) of 14 gates in 13 time steps give a CNOT, up to local unitary operations~\cite{Makhlin:2002p243}. Together with the additional 4 local unitary operations (labelled by $t_{1,2,3,4}$) shown in the outer block (in light blue), an exact CNOT gate can be implemented by a total of 18 gates in 17 time steps. Each coupling between the $i^{th}$ and $j^{th}$ qubits represents the unitary evolution $U_{ij}(t) = \mathrm{exp}\left(-\frac{i}{\hbar} t J \vec{\mathbf{S}}_i \cdot \vec{\mathbf{S}}_j  \right)$. Gate times are in units of $h/J$, so that gate times of $\frac{1}{4}$ and $\half$ correspond to $\sqrt{\textrm{SWAP}}$ and SWAP operations respectively.
}
\end{figure}
\addtocounter{subfigure}{1}

\section{Details of the experimental measurements}
This section presents the details of the experimental measurements shown in
Figs.~2 and~3 in the main text.
The data in Fig.~2 and 3 correspond to a single electron tunneling between the left
dot and the left reservoir, as shown in the inset to Fig.~2(a).  Charge sensing is performed
by measuring the current \sym{I}{QPC} using a bias voltage of \amount{500}{\mu V}.
We measure the electron loading and unloading rates (Fig.~2(a,b) in the main text) by applying a small amplitude square wave pulse to one of the gates defining the quantum dot and observing the electron tunneling events, which are manifested as steps in \sym{I}{QPC}. By averaging 200 of these single shot traces, we obtain a curve corresponding to the number of tunneling events per unit time interval.  The curves show an exponential decay (increase) for electron loading (unloading) events. Fitting these curves to exponentials allows the extraction of the tunnel rates for electron loading and unloading. To study the singlet and triplet states independently, we apply an in-plane magnetic field $B$.  As shown in Fig.~2(a), at \amount{B=3}{T}, where the triplet \Tm\ is the ground state\cite{Shi:2011unpublished}, we measure a loading rate of 521~Hz and an unloading rate of 645~Hz.  At \amount{B=1}{T}, shown in Fig.~2(b) and where the singlet is the ground state, we measure a loading rate of 81Hz and an unloading rate of 182Hz.

The difference in the unloading rate of singlet S and triplet \Tm\ can be used to measure the triplet-singlet relaxation time at magnetic fields $B$ small enough that the singlet S is the ground state~\cite{Hanson:2005p719}. One can preferentially load the \Tm\ by bringing the state in resonance with the Fermi level; with this initial condition, one then measures the unloading rate of the dot, which will show an exponential decay as a function of waiting time.  If the unloading rate is fast, it indicates that the electron stays in the \Tm\ state through the loading portion of the pulse and leaves from the \Tm\ state; if the unloading rate is slow, it indicates that the electron decays to the singlet state during the waiting period and leaves from state S.

To perform this measurement, it is useful to find the loading voltage that corresponds to resonance with the \Tm\ state, by measuring the loading and unloading tunnel rates as a function of loading depth. As shown in Fig.~\ref{fig:spectroscopy}, the loading rate develops two peaks as we increase the loading depth of the pulse. We identify the two peaks as \Tm\ and \Tz\ states coming into resonance with the Fermi level. These states have fast tunnel rates because the triplet wavefunction is more strongly coupled to the lead. The splitting between the peaks, when converted to energy using a lever arm measured in \cite{Simmons:2011p156804}, is~144 \units{\mu}eV between triplet \Tz\ and {\Tm},
a value that is in good agreement with the Zeeman splitting at \amount{B=1.5}{T}, using the $g$-factor measured in \cite{Simmons:2011p156804}. The unloading rate, when the loading depth is such that the \Tm\ state is in resonance with the Fermi level, is fast when the loading time is short, and is slow when the loading time is long. This indicates that during a long loading period, the electron loads into state \Tm\ and then relaxes to state S.  As shown in Fig.~\ref{fig:spectroscopy}(c), for short loading times, the electron remains in state \Tm\ until it unloads.

 \begin{figure}[h]
\includegraphics{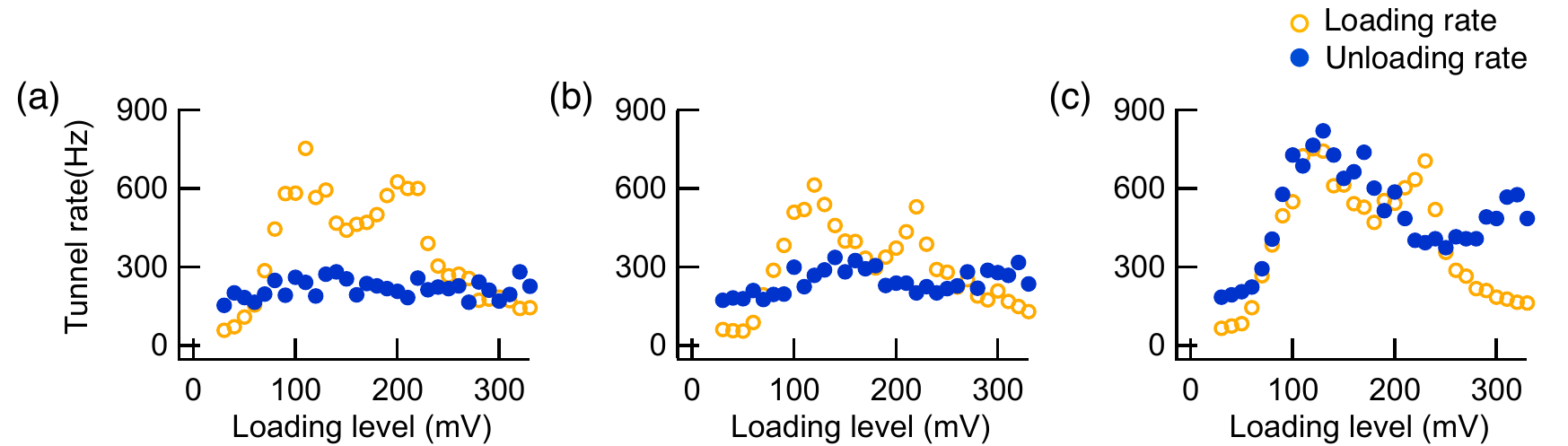}
\caption
{\label{fig:spectroscopy}Measurement of loading and unloading rates
into and out of the quantum dot as a function of loading depth. The unloading level is kept at -150mV. The unloading time is 40ms for all three panels and the loading times are \amount{600}{ms}, \amount{300}{ms} and \amount{40}{ms} for panel (a), (b), and (c), respectively. For long loading times, as in panel (a), the unloading rate is slow because the states \Tm\ and \Tz\ relax to the singlet state S during the loading period; for short loading times and loading voltages around \amount{100}{mV}, as in panel (c), the unloading rate is fast, indicating that the state \Tm\ does not relax to the singlet S before the electron tunnels out of the dot.}
\end{figure}
\addtocounter{subfigure}{1}  

Fig.~3 of the main text shows a measurement of the \Tm\ lifetime using this type of data.  The loading depth is set so that the state \Tm\ is in resonance with the Fermi level, and the unloading rate is measured as a function of loading time while keeping the unloading time constant (40ms). The unloading rate decays exponentially with loading time with a time constant of \amount{141}{ms} at \amount{B=1.5}{T}.

\end{onecolumngrid}

\bibliography{siliconQCsnc}

\end{document}